\documentclass[12pt]{iopart}

\usepackage{amssymb}
\usepackage[usenames,dvipsnames]{color} 
\usepackage{epsfig}

\usepackage{hyperref}

\hypersetup{
    colorlinks=true,
	linkcolor=Blue,
	citecolor=Blue,
	urlcolor=Blue
}

\newcommand{\muB}{\mu_\textrm{\footnotesize B}}
\newcommand{\rfig}[1]{Fig.~\ref{#1}}
\newcommand{\req}[1]{Eq.~\ref{#1}}
\newcommand{\rtab}[1]{Table~\ref{#1}}
\renewcommand{\emph}{\textit}

\begin{document}

\title{Buffer gas loaded magneto-optical traps for Yb, Tm, Er and Ho}

\author{Boerge Hemmerling$^{1,2}$, Garrett K.~Drayna$^{1,2,3}$, Eunmi Chae$^{1,2}$, Aakash Ravi$^{1,2}$ and John M.~Doyle$^{1,2}$}

\ead{boerge@cua.harvard.edu}

\address{$^1$ Harvard-MIT Center for Ultracold Atoms, Cambridge, Massachusetts 02138, USA}
\address{$^2$ Department of Physics, Harvard University, Cambridge, Massachusetts 02138, USA}
\address{$^3$ Department of Chemistry, Harvard University, Cambridge, Massachusetts 02138, MA}


\begin{abstract}
Direct loading of lanthanide atoms into magneto-optical traps (MOTs) from a very slow cryogenic buffer gas beam source is achieved, without the need for laser slowing. The beam source has an average forward velocity of 60-70\,m/s and a velocity half-width of $\sim 35$\,m/s, which allows for direct MOT loading of Yb, Tm, Er and Ho. Residual helium background gas originating from the beam results in a maximum trap lifetime of about 80\,ms (with Yb). The addition of a single-frequency slowing laser applied to the Yb in the buffer gas beam increases the number of trapped Yb atoms to $1.3(0.7) \times 10^8$ with a loading rate of $2.0(1.0) \times 10^{10}$\,atoms/s. Decay to metastable states is observed for all trapped species and decay rates are measured. Extension of this approach to the loading of molecules into a MOT is discussed.
\end{abstract}


\maketitle

\tableofcontents

\section{Introduction}

The invention of laser and evaporative cooling of atoms has led to significant new research, including new frequency standards, precision spectroscopy \cite{Riehle:03}, quantum computation \cite{Blatt:08, Monroe:02}, studies of cold collisions \cite{Julienne:99} and new degenerate quantum gases \cite{Ketterle:02}. The first laser cooling experiments were done with highly ``closed cycle'' species (alkali and alkaline-earth atoms), that show little spontaneous decay into metastable states. In subsequent work, species with more complex internal structure were cooled and trapped, including those with significant ``leaks'' into off-resonant states. Notable examples include Yb \cite{Yabuzaki:99}, Cr \cite{Celotta:00, Pfau:05}, Tm \cite{Sorokin:10}, Er \cite{Hanssen:06, Ferlaino:12} and Dy \cite{Lev:10}. The latter four have large magnetic dipole moments, suitable for the study of systems with long-range dipole-dipole interactions \cite{Santos:05} and the testing of fundamental theories \cite{Lev:10}, whereas Yb has been used in quantum simulation studies \cite{Takahashi:11b, Takahashi:11, Ueda:09}. Other possible uses for ultracold non $S$-state atoms could include the creation of exotic quantum phases and quantum magnetism \cite{Takahashi:12}.

Quantum many-body and quantum information research has driven the search for new species, including polar molecules. The additional degrees of freedom in molecules offer physical effects not found in atoms. Ultracold diatomic polar molecules are proposed candidates for novel quantum information and simulation experiments \cite{Zoller:06, DeMille:02} and new ultracold chemistry \cite{Ye:09, Friedrich:09}. The specific molecule necessary for an experiment can depend strongly on the desired physics to be studied. A method to provide a variety of molecules at mK temperatures and at a high phase-space density is desired, but is unavailable at present. 

One possible approach to useful sources of some (ultra-)cold molecules is laser cooling, including magneto-optical trapping. Recent experiments employing hydrodynamically enhanced cryogenic buffer gas beam sources (CBGBs) \cite{Doyle:12, Doyle:07} have reported laser cooling of molecules that have highly diagonal Frank-Condon factors: the optical slowing and cooling of SrF \cite{DeMille:10, DeMille:12}; the creation of a two-dimensional magneto-optical trap for YO \cite{Ye:13}; and the laser slowing of CaF \cite{Sauer:13} in a supersonic jet. Work is ongoing in several groups to create MOTs for these molecules.

In this paper, we report the successful use of a two-stage, slower CBGB for loading MOTs, including species that have leaky optical cycling transitions. This is a step toward a simple approach to loading polar diatomic molecules into a MOT. We demonstrate the method with lanthanide atoms. Employing a two-stage helium CBGB \cite{Doyle:11, Doyle:07}, we create cold, slow atomic beams of lanthanide atoms and load them directly into a magneto-optical trap. The effects of the collisions between residual helium from the CBGB and the trapped atoms are characterized. The feasibility of loading a MOT for molecules using this source is studied. The low mean forward velocity of our beam source is $\sim$\,65\,m/s, which renders a Zeeman slower unnecessary and thus allows for a direct loading of the MOT. This demonstrates a possible experimental path to a MOT for molecules; this beam is slower than the hydrodynamic CBGB used in other laser cooling experiments and thus offers a much shorter slowing path for the molecules, before the MOT. The initial cooling stage of our experiment relies only on collisions with He atoms (and thus is not dependent on the internal structure of the species).

\begin{figure}[th]
\centering
\includegraphics[width=0.95\columnwidth]{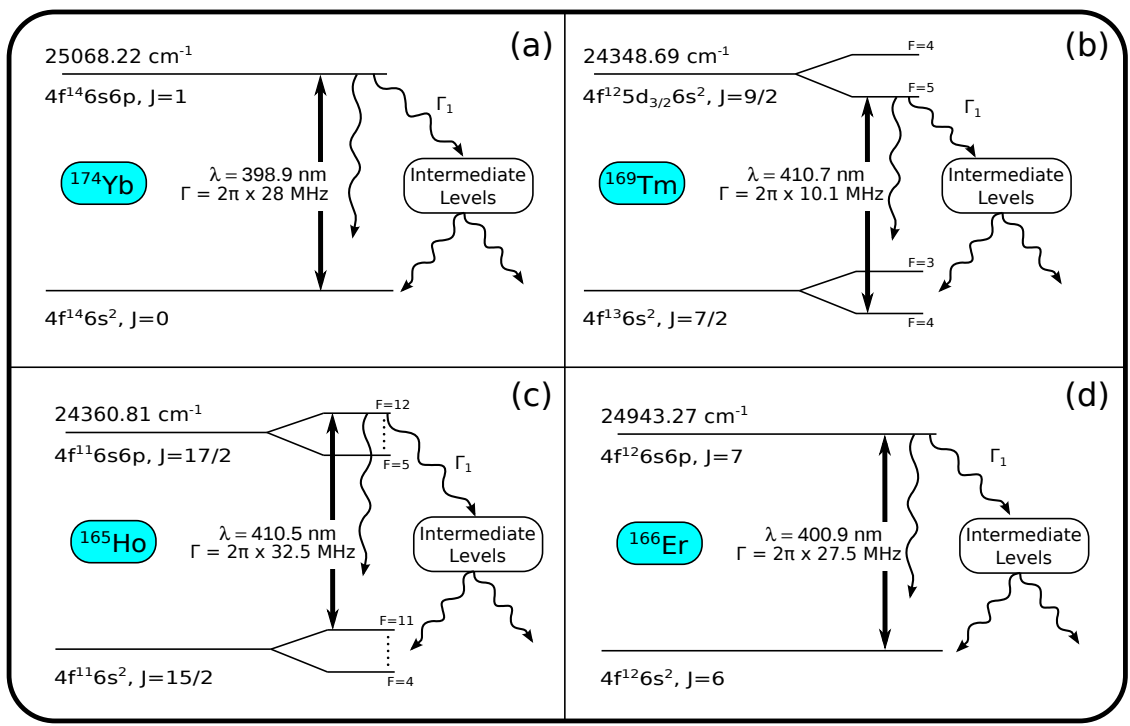}
\caption{Partial level scheme of all species with their corresponding laser cooling transition. Only the decay to metastable states ($\Gamma_1$) is considered for modeling the detuning dependent decay (\req{eq:lifetime_detuning_function}). The values are taken from \cite{Nist, Nave:03, Childs:83}.}
\label{fig:ho_level_scheme}
\end{figure}

We demonstrate a unique flexibility of this system with regard to the choice of species by creating MOTs for the elements Yb, Tm, Er and Ho in the same apparatus with no hardware change except tuning the MOT lasers, and no additional slowing light.

In the case of Yb, we load MOTs for the isotopes of mass 170, 171, 172, 174 and 176 by changing one parameter only, the detuning of the MOT laser frequency. The total number of atoms in common MOTs for Yb using the 400\,nm transition is limited to $\sim 4-5\times10^6$ \cite{Loftus:00, Yoon:03, Yabuzaki:03} due to spontaneous decay into metastable states during the loading phases of these traps ($\sim 1 - 10$\,s). We overcome this limitation with our high instantaneous flux beam source, allowing us to fully load the MOT in only a few milliseconds.

\section{Description of Atomic Species}

To most simply explain our method, we focus the discussion on only one species, Ho. Miao et al.~\cite{Saffman:13} have recently reported laser cooling and trapping of Ho. We also refer the reader to work on MOTs of the other species: Yb \cite{Yabuzaki:99}, Tm \cite{Sorokin:10} and Er \cite{Hanssen:06,Ferlaino:12}.

Ho has only one stable nuclear isotope, $^{165}$Ho. With its nuclear spin quantum number of $I=7/2$, it has the largest number of hyperfine states of any element, which makes it potentially useful for qubit implementations \cite{Moelmer:08}. Like many other lanthanides, it has a large magnetic dipole moment ($9\,\muB$). As shown in the level scheme in \rfig{fig:ho_level_scheme}(c), the transition at 410.5\,nm with a linewidth of $2\pi\times 32.5$\,MHz \cite{Nave:03, Childs:83} connecting the $4f^{11}6s^2 ( J=15/2, F=11)$ and $4f^{11}6s6p (J=17/2, F=12)$ states was used for laser cooling. As with previous work on other lanthanides (see e.g.~Ref.~\cite{Sorokin:10}), no additional repumper was necessary to trap Ho atoms in our MOT, despite the existence of possible decay channels into other hyperfine levels of the ground state manifold.

\section{Experimental Method}

\begin{figure}[t]
\begin{center}
\includegraphics[width=0.6\columnwidth]{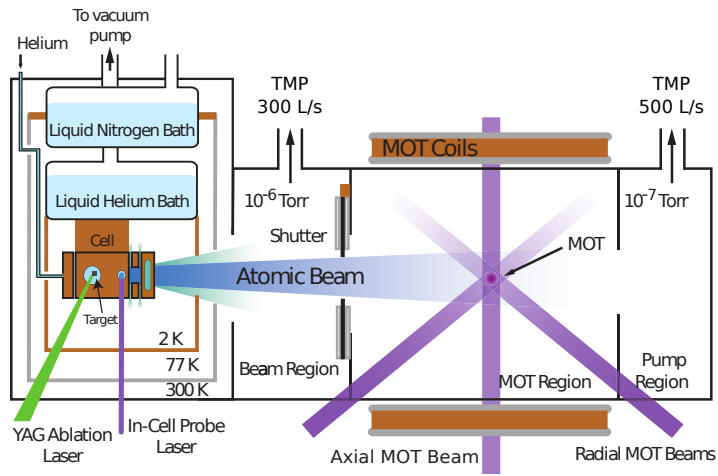}
\caption{\label{fig:apparatus}Schematic of the experimental apparatus. A cryogenic buffer gas beam source (left) opens into a room-temperature MOT section, separated by a differential pumping region. The photomultiplier tube and imaging system, which are located transversely to the atomic beam, are not depicted. Also not depicted is a 60\,L/s turbomolecular pump (TMP) on the MOT region. The distance between the exit of the cell and the MOT is 42\,cm.}
\end{center}
\end{figure}

The apparatus is shown in \rfig{fig:apparatus}. A detailed study of the two-stage buffer gas beam source is described elsewhere \cite{Doyle:11, Doyle:12, HsinI:13}. Briefly, the cell, operating at 2.5\,K, uses a combination of hydrodynamic extraction and a second slowing stage to produce a cold and effusive-like beam with a peak forward velocity of $\sim 60-70$\,m/s and FWHM of $\sim 70$\,m/s. A considerable fraction of atoms move below the capture velocity of the atomic MOTs, which is estimated to be $\lesssim 30$\,m/s for all species we trap here.

Atoms are introduced into the gas phase by laser ablation of solid precursor targets with 4\,ns long pulses and $\sim 14$\,mJ energy from a 532\,nm Nd:YAG laser. The atoms thermalize translationally via collisions with the cold He buffer gas inside the cell to a temperature of around 2 to 4\,K and leave the cell to form an atomic beam. We use He flow rates between 0.2-4\,sccm (standard cubic cm per minute). The beam travels from the cryogenic section, through a room temperature beam region, past a mechanical shutter, and into the MOT section. Since the atoms to be trapped are present only for a few ms after the ablation pulse, the shutter keeps residual He buffer gas from entering the MOT region when the atomic species of interest is not present in the beam. For the MOT measurements reported here, the shutter is opened for 10\,ms (with 3\,ms delay after ablation) during the MOT loading phase. In order to keep the amount of residual buffer gas background to a minimum, each section of the chamber is pumped by a turbomolecular pump to maintain steady-state low pressures of $\sim 1\times 10^{-6}$\,Torr in the beam region and $\sim 2\times 10^{-7}$\,Torr in the MOT regions.

We detect atoms in the MOT by imaging fluorescence onto a calibrated photomultiplier tube (PMT). To reduce background due to scattered laser light, we spatially filter the collected fluorescence using an objective that focuses the MOT image through a variable intermediate aperture with a minimal diameter of $\sim 600\mu$m. Simulation of the imaging system using commercial raytracing software, determines the collection efficiency to be $(2.5\pm0.5)\times 10^{-3}$.

MOT laser beams for each atomic species are derived from the same frequency-doubled Ti:Sapphire ring laser. Each MOT beam has $\sim 15$\,mW power with a $1/e^2$ diameter of $9.8 \pm 0.5$\,mm. The Ti:Saph laser is locked to a HeNe laser via a transfer cavity, providing $<5$\,MHz laser linewidth, which is used to determine the error in the lifetime measurements. The quadrupole field for the MOT is generated from a pair of water-cooled coils. The coils produce an axial (radial) field gradient of $3.8 \times 10^{-3}$\,T\,cm$^{-1}$  ($1.9 \times 10^{-3}$\,T\,cm$^{-1}$) for the Yb MOT and $1.7 \times 10^{-3}$\,T\,cm$^{-1}$ ($0.8 \times 10^{-3}$\,T\,cm$^{-1}$) for the Tm, Er and Ho MOTs, which is estimated by a finite-element software package.

\section{Results}

\subsection{Model for the Pulsed Loading Process}

The loading process of the MOT can be described by a phenomenological differential equation for the number of trapped particles $n(t)$ \cite{Foot:92},
\begin{eqnarray}
\label{eq:loading_dgl}
\frac{dn}{dt} = R(t) - \alpha n(t) - \beta n(t)^2\quad,
\end{eqnarray}
where $R(t)$ is the loading rate, $\alpha$ the loss due to background gas collisions and $\beta$ the intra-particle two-body loss. In our measurements, no evidence for two-body effects is observed and, thus, the corresponding term will henceforth be neglected ($\beta = 0$). The loading rate is time-dependent due to the pulsed nature of the loading process. We find that approximating the loading pulse by a Gaussian function as
\begin{eqnarray}
\label{eq:pulsed_load_func}
R(t) = \frac{n_\textrm{\footnotesize tot}}{\sqrt{2\pi} w} \cdot e^{-\frac{(t-t_0)^2}{2 w^2}}
\end{eqnarray}
yields a very good agreement with our measured data. The total number of atoms $n_\textrm{\footnotesize tot}$ is defined by the normalization $\int_{-\infty}^{\infty} dt R(t) = n_\textrm{\footnotesize tot}$, where $t_0$ is the pulse arrival time and $w$ the pulse width. The solution to the loading equation is
\begin{eqnarray}
\label{eq:pulsed_load_fit}
\nonumber
n(t) &=& \frac{1}{2}\cdot n_\textrm{\footnotesize tot} \cdot e^{\frac{1}{2} \alpha \left(-2(t-t_0) + \alpha w^2\right) }
\cdot\\*
&&\left(
\textrm{erf}\left(\frac{t_0+\alpha w^2}{\sqrt{2}w}\right)
-
\textrm{erf}\left(\frac{-t + t_0+\alpha w^2}{\sqrt{2}w}\right)
\right)\quad,
\end{eqnarray}
with the error function defined as $\textrm{erf}(x) = \frac{2}{\sqrt{\pi}} \int_0^x dt e^{-t^2}$.

\begin{figure}[t]
\centering
\includegraphics[width=0.9\columnwidth]{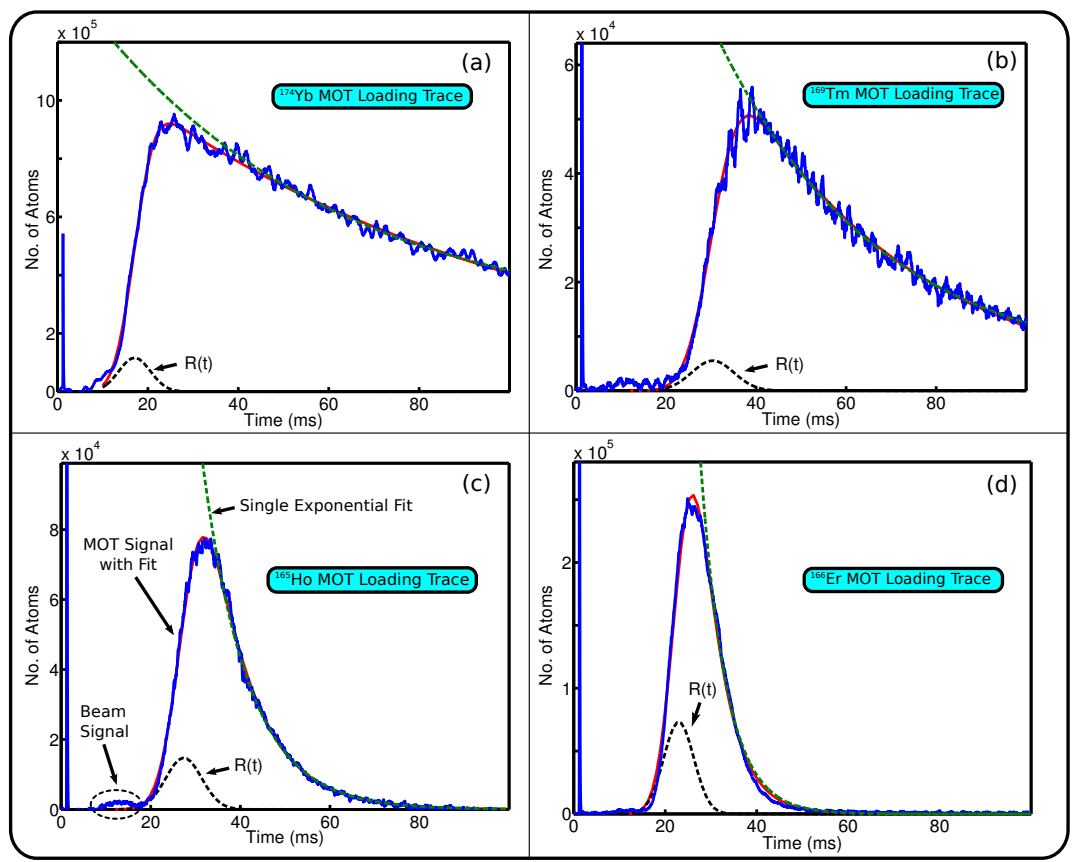}
\caption{Example time traces of the MOT loading process for each species with a He flow rate of 0.3\,sccm and a repetition rate of the ablation laser of 1.1\,Hz. The fit to \req{eq:pulsed_load_fit} agrees well with the data. The resulting fit parameters can be found in \rtab{tab:decay_to_metastable_states}. Also shown is the loading rate $R(t)$ corresponding to the fit and a single exponential fit for comparison.
}
\label{fig:ho_time_trace}
\end{figure}

Example of the measured time traces of the MOT loading process are shown in \rfig{fig:ho_time_trace} along with a fit to the pulsed loading model of \req{eq:pulsed_load_fit}. We focus on the case of Ho, \rfig{fig:ho_time_trace}\,(c). Note that only the tail of the buffer gas beam signal between 20-30\,ms is loaded into the MOT. While the beam signal represents only a velocity class corresponding to a certain detuning, the much larger MOT signal stems from particles from a range of velocities which have been actively cooled and trapped.
\begin{table}[b]
\caption{\label{tab:decay_to_metastable_states}MOT lifetimes, decay rates, saturation parameters and peak number of atoms from the fit to \req{eq:lifetime_detuning_function} for every species at 0.3\,sccm He flow rate and a repetition rate of the ablation laser of 1.1\,Hz (except for the last two rows for Yb). Both decay rates $\Gamma_0$ and $\Gamma_1$ are determined via extrapolation to zero He flow rate.}
\begin{indented}
\item[]\begin{tabular}{@{}lllll}
\br
& $^{174}$Yb & $^{169}$Tm & $^{165}$Ho & $^{166}$Er\\
\mr
$\Gamma_0$ (s$^{-1}$) & $10(2)$ & $20(2)$ & $38(7)$ & $113(16)$\\
$\Gamma_1$ (s$^{-1}$) & $9(^{+15}_{-9})^a$ & $12(^{+42}_{-12})^b$ & $1510(203)$ & $2071(753)^c$\\
1/$\alpha$ (ms) & $80(4)$ & $41(2)$ & $13(^{+18}_{-13})$ & $5.7(0.4)$\\
$s_{0,\textrm{\footnotesize eff}}$ & 1.65 & 1.60 & 0.40 & 0.47 \\
$N_\textrm{max}$ & $9.2 (4.4) \times 10^5$ & $5.0(2.7) \times 10^4$ & $7.7(3.1) \times 10^4$ & $2.5(1.6) \times 10^5$\\
$N_\textrm{max}$ & $9.0(4.7) \times 10^6{^d}$\\
$N_\textrm{max}^\textrm{slow}$ & $1.3(0.7) \times 10^8{^{d,f}}$\\
\br
\end{tabular}
\item[] $^a$ Previous work: 23(11)\,s$^{-1}$ from Ref.~\cite{Loftus:00}.
\item[] $^b$ Previous work: 22(6)\,s$^{-1}$ from Ref.~\cite{Sorokin:10}.
\item[] $^c$ Previous work: 1695(43)\,s$^{-1}$ from Ref.~\cite{Hanssen:06}.
\item[] $^d$ These measurements were taken with a repetition rate of the ablation laser of $0.6$\,Hz.
\item[] $^f$ This measurement employs a longitudinal single-frequency slowing laser.
\end{indented}
\end{table}
The resulting peak values for the observed number of atoms loaded into the MOT are given for each element in \rtab{tab:decay_to_metastable_states}. Note that the variation among the different species is not only caused by the cooling efficiency given by their corresponding natural linewidths of the cooling transition, but also by the ablation yields, which strongly vary between the specific atoms. The number of trapped atoms can be increased by applying low repetition rates of the ablation laser, leading to smaller heat loads on the buffer gas cell and more atoms below the capture velocity, and by using a single-frequency slowing laser. We demonstrate this with one of the elements, Yb; we observe a total of $1.3(0.7) \times 10^8$ trapped Yb atoms with an average loading rate of $n_\textrm{\footnotesize tot}/(2w) = 2.0(1.0) \times 10^{10}$\,atoms/s.

\begin{figure}[t]
\centering
\includegraphics[width=0.6\columnwidth]{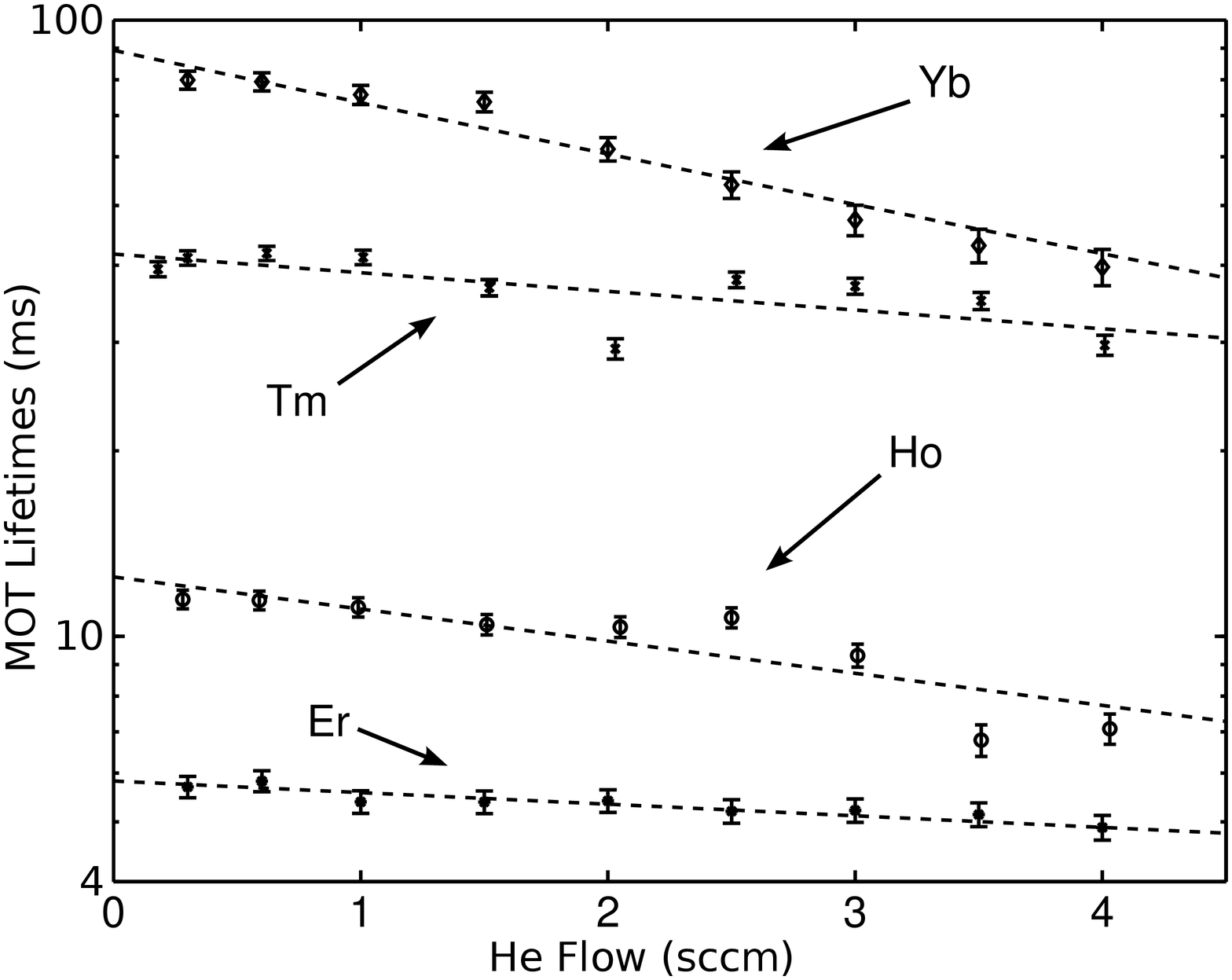}
\caption{MOT lifetimes ($1/\alpha$) for every species from fitting the time traces to \req{eq:pulsed_load_fit}. The dashed lines are exponential fits and act as a guide to the eye only.}
\label{fig:lifetime_vs_he_flow}
\end{figure}

\subsection{Limitation of the MOT Lifetime due to the Helium Buffer Gas}

Two processes compete with the trap loading and limit the MOT lifetime, namely, collisions with the background gas and decay into metastable states. The first process is dominated by residual He buffer gas in the MOT region. After cold He hits the room-temperature walls of the vacuum chamber, it can bounce back and hit the trapped atoms, leading to losses. In the case of Yb, for example, a head-on collision of an atom at rest with a room-temperature He atom increases its speed by $\sim 50$\,m/s, which means the atom cannot be recaptured into the trap. The measured decrease of MOT lifetimes with varying He flow is shown in \rfig{fig:lifetime_vs_he_flow}. All measurements were taken at the laser detuning which results in the maximum MOT fluorescence signal. Each point is the result of averaging over five ablation shots. The highest lifetimes for each element were observed for $\sim 0.3$\,sccm He flow and are listed in \rtab{tab:decay_to_metastable_states}.

\subsection{Decay into Metastable States}

The large differences among the species are due to the second lifetime-limiting effect, namely the presence of decay channels to metastable states. Once the atom decays after a number of photon scattering events into such a dark state, it escapes the cooling and trapping process. In order to fully describe this decay, for example in the case of Er, 110 intermediate states have to be considered \cite{Hanssen:06}. As a result, theoretical modeling of such complex atoms is daunting. Instead, we apply a simplified model \cite{Sorokin:10, Hanssen:06, Loftus:00} which considers a three-level system including trap losses (\rfig{fig:ho_level_scheme}). In this model, the lifetime $1/\alpha$ scales with the fractional population in the excited state and depends on the laser detuning as
\begin{eqnarray}
\label{eq:lifetime_detuning_function}
\alpha(\Delta) = \Gamma_0 + \Gamma_1 \cdot \left( \frac{s_{0, \textrm{\footnotesize eff}}/2}{1+s_{0, \textrm{\footnotesize eff}}+4\Delta^2/\Gamma^2} \right)\quad,
\end{eqnarray}
where $\Gamma_0$ represents the loss rate due to background collisions, $\Gamma_1$ is the decay rate into intermediate metastable states, $s_{0, \textrm{\footnotesize eff}} = \kappa \cdot s_0$ is the effective saturation parameter, $\Delta$ is the laser detuning from the resonance, and $\Gamma = 2\pi\times \Delta\nu$, where $\Delta\nu$ is the natural linewidth. The effective saturation parameter takes averaging over Zeeman substates and random light polarization in the MOT region into account, where $\kappa \approx 3 \cdot \frac{2 F + 1}{2 F + 3}$ and $F$ corresponds to the hyperfine quantum number of the ground state. This simplified model omits any population which gets recycled back into the ground state after the decay to the metastable state reservoir, but allows for the determination of a lower limit of the decay rate into the intermediate states.

\begin{figure}[t]
\centering
\includegraphics[width=0.9\columnwidth]{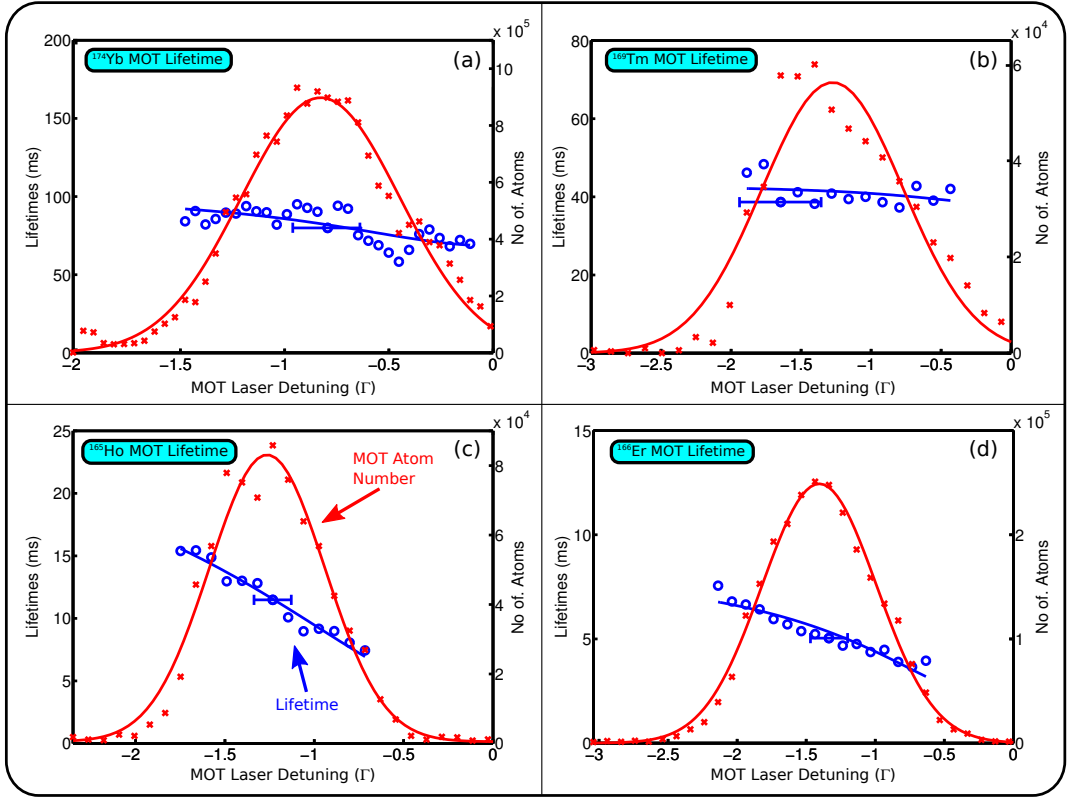}
\caption{MOT lifetimes ($1/\alpha$) of all species as function of the laser detuning in units of the natural linewidth with a He flow rate of 0.3\,sccm and a repetition rate of the ablation laser of 1.1\,Hz. Negative values are red-detuned with respect to the line center. The solid line is a fit to \req{eq:lifetime_detuning_function}. Also shown is the relative atom number in the MOT, including a Gaussian fit as a guide to the eye.}
\label{fig:lifetime_vs_detuning}
\end{figure}

The measurements for each species are shown in \rfig{fig:lifetime_vs_detuning}. Again, we focus on the case of Ho, \rfig{fig:lifetime_vs_detuning}\,(c), as an example. We observe the largest MOT fluorescence signal with a red-detuned laser at $\approx -1.3\,\Gamma$. Although this is the optimal detuning for creating high trapped atom numbers, the longest MOT lifetime is observed at detunings further to the red. This effect is explained by the lower population in the excited state and, consequently, a lower probability of decaying into a metastable dark state. Increasing the excited state population by moving the laser frequency closer to the transition line center decreases the MOT lifetime almost by a factor of three. The strength of this effect is governed by the decay rate $\Gamma_1$. The measurement results obtained for each atomic species are summarized in \rtab{tab:decay_to_metastable_states}. For the fitting, $\Gamma_{0}$ and $\Gamma_{1}$ were free parameters. Our measurements indicate an increase of the decay rates with an increase of the He flow rate for all species. In the table, we quote values of the decay rates which are extrapolated to zero He flow.

The collision rate with the background gas $\Gamma_0$ can be estimated as $\Gamma^{est}_0 = n_\textrm{\tiny He} \cdot \bar{v}_\textrm{\tiny He} \cdot \sigma_\textrm{\tiny He-Atom}$ \cite{Kirk:09}, where $n_\textrm{\tiny He} = \frac{P}{k_\textrm{\tiny B} \cdot T}$ is the He gas density and $k_\textrm{\tiny B}$ is the Boltzmann constant. We assume that He at room temperature ($T = 300\,$K) is the dominant part of the background gas. With a cross-section of $\sigma_\textrm{\tiny He-Atom} \approx 10^{-14}$\,cm$^2$, the velocity $\bar{v}_\textrm{\tiny He} \approx 1120$\,m/s , we find that $\Gamma^{est}_0$ ranges from 4 to 36\,s$^{-1}$ for steady-state pressures between $P = 10^{-7}$ and $10^{-6}$\,Torr, which is consistent with our measurements.

The results for $\Gamma_1$ for Yb, Tm and Er are in agreement within the errors with previously reported work. For Ho, both our measurement and the result of the Saffman group \cite{Saffman:13} are currently the only experimental estimates of the decay rate to metastable states. At present, no theoretical prediction is available.

\section{Conclusion}

In conclusion, we demonstrated, using Yb, Tm, Er and Ho, the direct loading of magneto-optical traps from a cryogenic two-stage He buffer gas beam source. Despite the presence of the He background gas, we observed lifetimes of up to 80\,ms (for Yb). We report a decay rate to metastable states for Ho of $1510(203)$\,s$^{-1}$. 

For further applications, we envision an experiment where the buffer gas source and the trapping region are separated by a differential pumping stage to further reduce the residual He background pressure near the MOT. Moreover, for atoms or molecules with magnetic ground states, a magnetic guide \cite{Doyle:07} could aid in this process by redirecting only the species of interest into a ultra-high vacuum region. 

The unprecedentedly high loading rate of $2.0(1.0) \times 10^{10}$\,atoms/s, which we demonstrated with Yb, will allow for loading a high number of atoms with loss mechanisms that have a time-scale comparable to the trap loading time into a consecutive trap. As an example, the number of atoms in a blue Yb MOT is limited by spontaneous decay into metastable states. This is typically overcome by using an intermediate MOT with a narrow intercombination line \cite{Yabuzaki:03} which has no optical leaks. As an alternative, our approach would surpass the need of such an intermediate step and simplify experimental requirements. If developed further, it could produce quantum gases with larger atom number, at a higher repetition rate. This is particularly enticing as the cryogenic setup used is very straightforward.

The great flexibility and low intrinsic temperature of this atomic and molecular beam source make it a possible tool for implementing MOTs for any species that precludes the common approach of a combination of a high-temperature oven and Zeeman slower. In particular, recent experiments on slowing and cooling diatomic molecules \cite{DeMille:10, DeMille:12, Ye:13}, all of which successfully employ buffer gas beam sources, reflect this potential advantage.

Moreover, our approach constitutes a straightforward testbed for laser cooling and trapping of other atomic species which have not been studied yet. In particular, elements with low vapor pressures at temperatures which are feasible with common oven sources are interesting candidates since laser ablation can produce high densities of such elements in the gas phase. Since the initial cooling stage of our beam source relies solely on collisions with He, only a single laser providing the optical force of the magneto-optical trap has to be adapted to the wavelength of the corresponding species. At the same time, one can imagine the simultaneous loading of multiple species out of the same source. This is a major advantage over regular setups where a separate Zeeman slower is needed for each species loaded into a magneto-optical trap.

\ack

We would like to thank Jun Ye, Mark Yeo, Matthew Hummon, Alejandra Collopy, Benjamin Stuhl, Hsin-I Lu and Nicholas R.~Hutzler for helpful discussions. We also acknowledge the contributions of Matthew J.~Wright to the initial phase of this experiment. This work was supported by the NSF and the ARO. We thank Mark Saffman for helpful discussions in the preparation of this manuscript.

\section*{References}

\bibliography{Hemmerling_Buffer_Gas_Loaded_MOTs_for_Yb_Tm_Er_and_Ho_literature}

\end{document}